\documentclass[apjl]{emulateapj}
\usepackage{graphicx}
\usepackage{amssymb}

\usepackage{natbib}

\begin{document}
\title{Discovery of TeV Gamma Ray Emission from Tycho's Supernova Remnant}
\author{
V.~A.~Acciari\altaffilmark{1},
E.~Aliu\altaffilmark{2},
T.~Arlen\altaffilmark{3},
T.~Aune\altaffilmark{4},
M.~Beilicke\altaffilmark{5},
W.~Benbow\altaffilmark{1},
S.~M.~Bradbury\altaffilmark{6},
J.~H.~Buckley\altaffilmark{5},
V.~Bugaev\altaffilmark{5},
K.~Byrum\altaffilmark{7},
A.~Cannon\altaffilmark{8},
A.~Cesarini\altaffilmark{9},
L.~Ciupik\altaffilmark{10},
E.~Collins-Hughes\altaffilmark{8},
W.~Cui\altaffilmark{11},
R.~Dickherber\altaffilmark{5},
C.~Duke\altaffilmark{12},
M.~Errando\altaffilmark{2},
J.~P.~Finley\altaffilmark{11},
G.~Finnegan\altaffilmark{13},
L.~Fortson\altaffilmark{10},
A.~Furniss\altaffilmark{4},
N.~Galante\altaffilmark{1},
D.~Gall\altaffilmark{11},
G.~H.~Gillanders\altaffilmark{9},
S.~Godambe\altaffilmark{13},
S.~Griffin\altaffilmark{14},
J.~Grube\altaffilmark{10},
R.~Guenette\altaffilmark{14},
G.~Gyuk\altaffilmark{10},
D.~Hanna\altaffilmark{14},
J.~Holder\altaffilmark{15},
J.~P.~Hughes\altaffilmark{16},
C.~M.~Hui\altaffilmark{13},
T.~B.~Humensky\altaffilmark{17},
P.~Kaaret\altaffilmark{18},
N.~Karlsson\altaffilmark{10},
M.~Kertzman\altaffilmark{19},
D.~Kieda\altaffilmark{13},
H.~Krawczynski\altaffilmark{5},
F.~Krennrich\altaffilmark{20},
M.~J.~Lang\altaffilmark{9},
S.~LeBohec\altaffilmark{13},
A.~S~Madhavan\altaffilmark{20},
G.~Maier\altaffilmark{21},
P.~Majumdar\altaffilmark{3},
S.~McArthur\altaffilmark{5},
A.~McCann\altaffilmark{14},
P.~Moriarty\altaffilmark{22},
R.~Mukherjee\altaffilmark{2},
R.~A.~Ong\altaffilmark{3},
M.~Orr\altaffilmark{20},
A.~N.~Otte\altaffilmark{4},
D.~Pandel\altaffilmark{18},
N.~H.~Park\altaffilmark{17},
J.~S.~Perkins\altaffilmark{1},
M.~Pohl\altaffilmark{23},
J.~Quinn\altaffilmark{8},
K.~Ragan\altaffilmark{14},
L.~C.~Reyes\altaffilmark{17},
P.~T.~Reynolds\altaffilmark{24},
E.~Roache\altaffilmark{1},
H.~J.~Rose\altaffilmark{6},
D.~B.~Saxon\altaffilmark{15,*},
M.~Schroedter\altaffilmark{20},
G.~H.~Sembroski\altaffilmark{11},
G.~Demet~Senturk\altaffilmark{25},
P.~Slane\altaffilmark{26},
A.~W.~Smith\altaffilmark{7},
G.~Te\v{s}i\'{c}\altaffilmark{14},
M.~Theiling\altaffilmark{1},
S.~Thibadeau\altaffilmark{5},
K.~Tsurusaki\altaffilmark{18},
A.~Varlotta\altaffilmark{11},
V.~V.~Vassiliev\altaffilmark{3},
S.~Vincent\altaffilmark{13},
M.~Vivier\altaffilmark{15},
S.~P.~Wakely\altaffilmark{17,**},
J.~E.~Ward\altaffilmark{8},
T.~C.~Weekes\altaffilmark{1},
A.~Weinstein\altaffilmark{3},
T.~Weisgarber\altaffilmark{17},
D.~A.~Williams\altaffilmark{4},
M.~Wood\altaffilmark{3},
B.~Zitzer\altaffilmark{11}
}

\altaffiltext{1}{Fred Lawrence Whipple Observatory, Harvard-Smithsonian Center for Astrophysics, Amado, AZ 85645, USA}
\altaffiltext{2}{Department of Physics and Astronomy, Barnard College, Columbia University, NY 10027, USA}
\altaffiltext{3}{Department of Physics and Astronomy, University of California, Los Angeles, CA 90095, USA}
\altaffiltext{4}{Santa Cruz Institute for Particle Physics and Department of Physics, University of California, Santa Cruz, CA 95064, USA}
\altaffiltext{5}{Department of Physics, Washington University, St. Louis, MO 63130, USA}
\altaffiltext{6}{School of Physics and Astronomy, University of Leeds, Leeds, LS2 9JT, UK}
\altaffiltext{7}{Argonne National Laboratory, 9700 S. Cass Avenue, Argonne, IL 60439, USA}
\altaffiltext{8}{School of Physics, University College Dublin, Belfield, Dublin 4, Ireland}
\altaffiltext{9}{School of Physics, National University of Ireland Galway, University Road, Galway, Ireland}
\altaffiltext{10}{Astronomy Department, Adler Planetarium and Astronomy Museum, Chicago, IL 60605, USA}
\altaffiltext{11}{Department of Physics, Purdue University, West Lafayette, IN 47907, USA }
\altaffiltext{12}{Department of Physics, Grinnell College, Grinnell, IA 50112-1690, USA}
\altaffiltext{13}{Department of Physics and Astronomy, University of Utah, Salt Lake City, UT 84112, USA}
\altaffiltext{14}{Physics Department, McGill University, Montreal, QC H3A 2T8, Canada}
\altaffiltext{15}{Department of Physics and Astronomy and the Bartol Research Institute, University of Delaware, Newark, DE 19716, USA}
\altaffiltext{16}{Department of Physics and Astronomy, Rutgers University, 136 Frelinghuysen Rd., Piscataway, NJ 08854-8019}
\altaffiltext{17}{Enrico Fermi Institute, University of Chicago, Chicago, IL 60637, USA}
\altaffiltext{18}{Department of Physics and Astronomy, University of Iowa, Van Allen Hall, Iowa City, IA 52242, USA}
\altaffiltext{19}{Department of Physics and Astronomy, DePauw University, Greencastle, IN 46135-0037, USA}
\altaffiltext{20}{Department of Physics and Astronomy, Iowa State University, Ames, IA 50011, USA}
\altaffiltext{21}{DESY, Platanenallee 6, 15738 Zeuthen, Germany}
\altaffiltext{22}{Department of Life and Physical Sciences, Galway-Mayo Institute of Technology, Dublin Road, Galway, Ireland}
\altaffiltext{23}{Institut f\"ur Physik und Astronomie, Universit\"at Potsdam, 14476 Potsdam-Golm,Germany}
\altaffiltext{24}{Department of Applied Physics and Instrumentation, Cork Institute of Technology, Bishopstown, Cork, Ireland}
\altaffiltext{25}{Columbia Astrophysics Laboratory, Columbia University, New York, NY 10027, USA}
\altaffiltext{26}{Harvard-Smithsonian Center for Astrophysics, 60 Garden Street, Cambridge, MA 02138, USA}

\altaffiltext{*}{Corresponding author: dbsaxon@udel.edu}
\altaffiltext{**}{Corresponding author: wakely@uchicago.edu}

\begin{abstract}

We report the discovery of TeV gamma-ray emission from the Type Ia supernova remnant (SNR) G120.1+1.4, known as Tycho's supernova remnant.  Observations performed in the period 2008-2010 with the VERITAS ground-based gamma-ray observatory reveal weak emission coming from the direction of the remnant, compatible with a point source located at $00^{\rm h} \ 25^{\rm m} \ 27.0^{\rm s},\ +64^{\circ} \ 10^{\prime} \ 50^{\prime\prime}$ (J2000).  The TeV photon spectrum measured by VERITAS can be described with a power-law $dN/dE = C(E/3.42\;\textrm{TeV})^{-\Gamma}$ with $\Gamma = 1.95 \pm 0.51_{stat} \pm 0.30_{sys}$ and $C = (1.55 \pm 0.43_{stat} \pm 0.47_{sys}) \times 10^{-14}$ cm$^{-2}$s$^{-1}$TeV$^{-1}$. The integral flux above 1 TeV corresponds to $\sim 0.9\%$ percent of the steady Crab Nebula emission above the same energy, making it one of the weakest sources yet detected in TeV gamma rays.  We present both leptonic and hadronic models which can describe the data.  The lowest magnetic field allowed in these models is $\sim 80 \mu$G, which may be interpreted as evidence for magnetic field amplification.
\end{abstract}

\keywords{gamma rays: observations --- ISM: individual objects (G120.1+01.4, Tycho=VER J0025+641)}

\section{Introduction}

The object G120.1+1.4, commonly called Tycho's supernova remnant (SNR), is the historical relic of a supernova which was first observed in 1572.  Recent spectral analysis of the light echo from the explosion \citep{2008Natur.456..617K} has confirmed the long-standing conjecture \citep[see, e.g.,][]{2004ApJ...612..357R} that the event was a Type Ia supernova.

At 438 years old, Tycho is young among Galactic SNRs and is well-studied at many wavelengths.  It has a radio spectral index of $0.65$ and a flux density at 1.4 GHz of $40.5$ Jy \citep{2006A&A...457.1081K}.  A hint of spectral curvature, consistent with nonlinear shock acceleration, and a slightly flatter radio spectrum (0.61) has also been reported \citep{1992ApJ...399L..75R}.  Radio images show a clear shell-like morphology with enhanced emission along the northeastern edge of the remnant \citep{1991AJ....101.2151D,2009ApJ...696.1864S}.

X-ray images reveal strong non-thermal emission concentrated in the SNR rim \citep{2002JApA...23...81H,2005ApJ...621..793B,2005ApJ...634..376W,2010ApJ...709.1387K}.  Thin filamentary X-ray structures in this region have been interpreted as evidence for electron acceleration \citep{2002JApA...23...81H,2005ApJ...621..793B,2005ApJ...634..376W}.  This is supported by radio spectral tomography studies \citep{2000ApJ...529..453K} and by the detection of X-rays up to energies of 30 keV \citep[e.g.,][]{2009PASJ...61S.167T}, which implies the presence of electrons up to at least $\sim10$ TeV.

The global expansion rate of the remnant, measured at many wavelengths \citep{1978ApJ...224..851K,1997ApJ...491..816R,2010ApJ...709.1387K}, is consistent with an object transitioning into the Sedov phase.  However, many of these studies have noted that the northeast quadrant of the remnant is expanding at a lower rate than the rest of the object.  This has been attributed by some to interactions with a nearby high-density cloud, which has been studied through its HI and CO emissions \citep{1999AJ....117.1827R,2004ApJ...605L.113L,2009ChA&A..33..393C}.

Distance estimates for Tycho have varied.  Estimates which combine measurements of proper motion with shock velocities inferred from H$_{\alpha}$ line widths tend to favor distances near $\sim2.5$~kpc \citep{1980ApJ...235..186C,1991ApJ...375..652S,1992ApJ...384..665S}.  On the other hand, the distance derived from the light echo spectrum \citep{2008Natur.456..617K} is $3.8^{+1.5}_{-1.1}$ kpc, and HI absorption studies \citep{1995A&A...299..193S} yield $4.5 \pm 0.5$~kpc.  A recent X-ray study combining ejecta velocities with proper motion studies yields $4.0 \pm 1.0$~kpc \citep{2010arXiv1009.6031H}.  The apparent association with a molecular CO cloud has not produced an unambiguous distance estimate \citep[e.g.,][]{2004ApJ...605L.113L,2009ChA&A..33..393C}, due to the complex velocity field in the direction of the Perseus arm, where Tycho appears to be located.  For the same reason, estimates based on the HI measurements may be considered controversial.  Finally, \cite{2008A&A...483..529V} derive a lower limit of 3.3~kpc, based on the previous non-detection of Tycho in TeV gamma rays.

It is worth noting that many of the shorter distance estimates rely on assumptions about the post-shock proton temperature's relationship to the shock velocity in the case of adiabatic expansion \citep[see also][]{1988MNRAS.230..331S}.  These assumptions may not be valid in the presence of efficient particle acceleration, as discussed, for instance, in \cite{2009Sci...325..719H}.


Claims of such efficient nuclear particle acceleration have indeed been made (e.g., \citealt{2005ApJ...634..376W}, though see also \citealt{2010ApJ...715L.146L}), based on detailed studies of the shock front and contact discontinuity locations.  Drawing on this result and on their own detailed simulations, \cite{2008A&A...483..529V} claim that a detection of gamma rays from Tycho would represent ``incontrovertible" evidence of nuclear particle acceleration in SNRs.  More generally, TeV gamma-ray observations probe the high-energy end of the underlying parent particle distributions and, especially in conjunction with measurements at other wavelengths, can help constrain the associated acceleration mechanics (see, e.g., \citealt{2009astro2010S.145K}).

Tycho has been observed several times at gamma-ray energies, with no credible detections previously reported.  In the GeV regime, the source does not appear in either the first Fermi LAT catalog \citep{2010ApJS..188..405A} or
the 3rd EGRET catalog \citep{1999ApJS..123...79H}.  Upper limits in the TeV range have been presented by the Whipple Collaboration \citep{1998A&A...329..639B}, the HEGRA Collaboration \citep{2001A&A...373..292A} and the MAGIC Collaboration \citep{2009arXiv0907.1009C}.  This last limit is the most constraining, with a $3\sigma$ (where $\sigma$ indicates standard deviations) flux upper limit of about 1.7\% of the steady Crab Nebula flux above 1 TeV.  All of the TeV limits assume a gamma-ray source located at the center of the remnant.

In this Letter we report the discovery by VERITAS of TeV gamma-ray emission from Tycho's supernova remnant.

\section{VERITAS Instrument \& Observations}

VERITAS is an array of four 12~m imaging atmospheric Cherenkov telescopes located at the Fred Lawrence Whipple Observatory in southern Arizona (1.3 km a.s.l., N~31$^\circ$40$^\prime$, W~110$^\circ$57$^\prime$).  Each of the telescopes has a camera comprising 499 photomultiplier tubes covering a $3.5^{\circ}$ total field of view.  The array is sensitive to photon energies between 100~GeV and 30~TeV, with an energy resolution of $\sim15\%$ and an angular resolution per event of $\sim0.1^{\circ}$.  The array has been fully functional since September 2007.  In summer 2009, telescope T1 was relocated, improving the sensitivity and angular resolution of the array \citep{2009arXiv0912.3841P}.  Prior to this reconfiguration, VERITAS was capable of detecting a point source with 1\% of the flux of the Crab Nebula at the $5\sigma$ level in less than 50 hours.  After moving T1, such a source can be detected in less than 30 hours.

VERITAS observations of Tycho spanned two epochs, from October 2008 to January 2009 and from September 2009 to January 2010.  All observations were taken in  ``wobble mode'', in which the telescopes are pointed to a position $0.5^{\circ}$ offset from the source position.  The offset direction is sequentially varied from run to run between the four cardinal directions. Approximately equal amounts of data were acquired from each offset position.  This technique allows for the collection of data with a simultaneous estimate of the background.  From the period between October 2008 and January 2009, a total of 21.9 hours of data were retained after quality cuts on weather conditions and operational stability.  The mean zenith angle of these data is $35^{\circ}$.  Between September 2009 and January 2010, an additional 44.7 hours of data were analyzed after quality cuts.  The mean zenith angle of this data set is $39^{\circ}$.  Note that the second season's data set was taken with the new array configuration, and therefore benefits from the improved sensitivity.

\section{Data Analysis and Results}

The data were analyzed following standard procedures \citep[see, e.g.,][]{2008ApJ...679.1427A}. The optimum gamma-ray selection criteria (\textit{cuts}) depend upon the source flux and spectrum, which are not known a priori. We therefore define and apply two sets of cuts, appropriate for a reasonable range of source properties, and account for the statistical trials incurred in this process. The results presented here required that at least 3 of the telescopes in the array recorded an image with more than 800 digital counts ($\sim150$ photo-electrons, corresponding to an energy threshold of ~800 GeV). These images were then used to reconstruct the air shower properties. Gamma-ray events were selected through cuts on the \emph{mean reduced scaled width} and \emph{mean reduced scaled length} parameters \citep{2006APh....25..380K}, which were required to fall between -1.2 and 0.5, and on the square of the angular distance from the test position to the reconstructed arrival direction of the shower, $\theta^2$.  For the 2008-2009 data set, we required $\theta^2 < 0.015\,\textrm{deg}^2$, and for the 2009-2010 data set, we required $\theta^2 < 0.01\,\textrm{deg}^2$ because of the improved angular resolution of the array.  The ring background model was used to estimate the background \citep[see, e.g.,][for a description]{2007A&A...466.1219B}.

This analysis produced an excess which is significant at the $5.8\sigma$ level (pre-trials). This is the peak excess found in a blind search region with sides of length 0.26$^{\circ}$ - roughly twice the diameter of the radio remnant.  A conservative a priori trials factor was determined by tiling this area with square 0.04$^{\circ}$ bins \citep{2006ApJ...636..777A}, and additionally accounting for the two sets of applied cuts.  This results in a $5.0\sigma$ post-trials significance.

\begin{figure}
\includegraphics[width=0.99\linewidth]{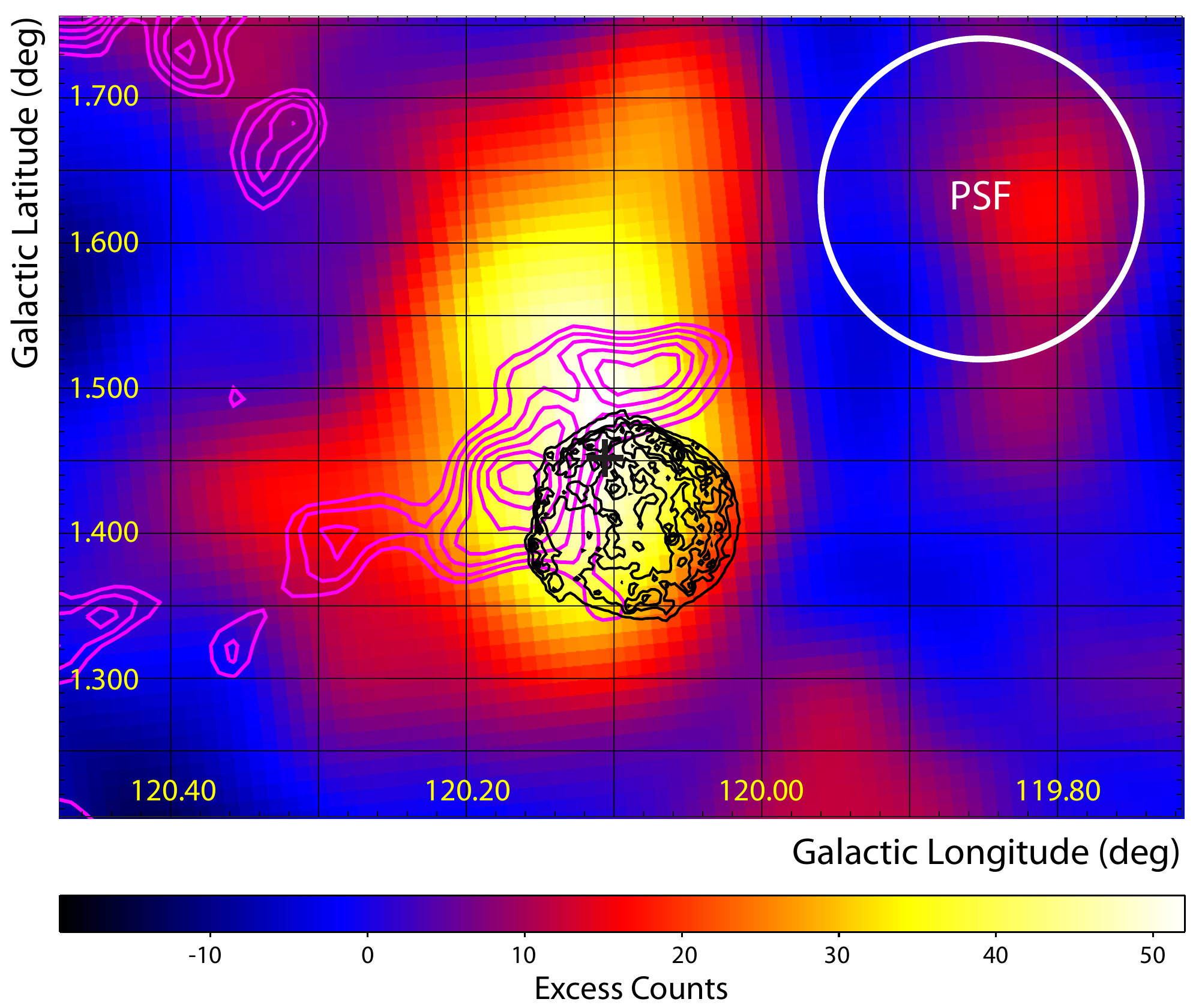}
\caption{\label{fig:skymap}VERITAS TeV gamma-ray count map of the region around Tycho's SNR.  The color scale indicates the number of excess gamma-ray events from a region, using a squared integration radius of 0.01~deg$^2$ for the 2009/2010 data and 0.015 deg$^2$ for the 2008/2009 data.  The centroid of the emission is indicated with a thick black cross.  Overlaid on the image are X-ray contours from a Chandra ACIS exposure \citep[thin black lines;][]{2002JApA...23...81H} and $^{12}$CO emission (J=1-0) from the high-resolution FCRAO Survey \citep[magenta lines;][]{1998ApJS..115..241H}.  The CO velocity selection is discussed in the text.  The VERITAS count map has been smoothed with Gaussian kernel of size $0.06^\circ$.  The point-spread function of the instrument (see text) is indicated by the white circle.}
\end{figure}

%

\subsection{Morphology}

The morphology of the source was investigated by binning the uncorrelated acceptance-corrected map of excess event counts.  Bins of size $0.05^\circ$ were used to provide sufficient statistics for fits to source models.  The map is compatible ($\chi^2=508;\textrm{ndf}=438;\textrm{Prob}=1.2\%$) with a point source located at $00^{\rm h} \ 25^{\rm m} \ 27.0^{\rm s},\ +64^{\circ} \ 10^{\prime} \ 50^{\prime\prime}$ (J2000) and hence we designate the object VER~J0025+641. This position is derived from a simple symmetric Gaussian fit with a width fixed at the instrument point-spread function (i.e., the 68\% containment radius for photons, $\theta_{68\%}$= 0.11$^{\circ}$).  While other source functions (e.g., an offset asymmetric Gaussian) may provide a marginally better fit (perhaps hinting at a more complex underlying source morphology) a likelihood ratio test shows that the extra degrees of freedom are not statistically justified in this data set.

As shown in Figure 1, the center of the fit position is offset by 0.04$^{\circ}$ from the center of the remnant. The statistical uncertainty in this location is $0.023^{\circ}$, while the systematic uncertainty resulting from telescope pointing accuracy is $0.014^{\circ}$. We note that the derived centroid position depends also on the source shape assumed for the fit.  Future observations will allow more detailed study of the source morphology.

\subsection{Spectrum}
\begin{figure}
\includegraphics[width=0.99 \linewidth]{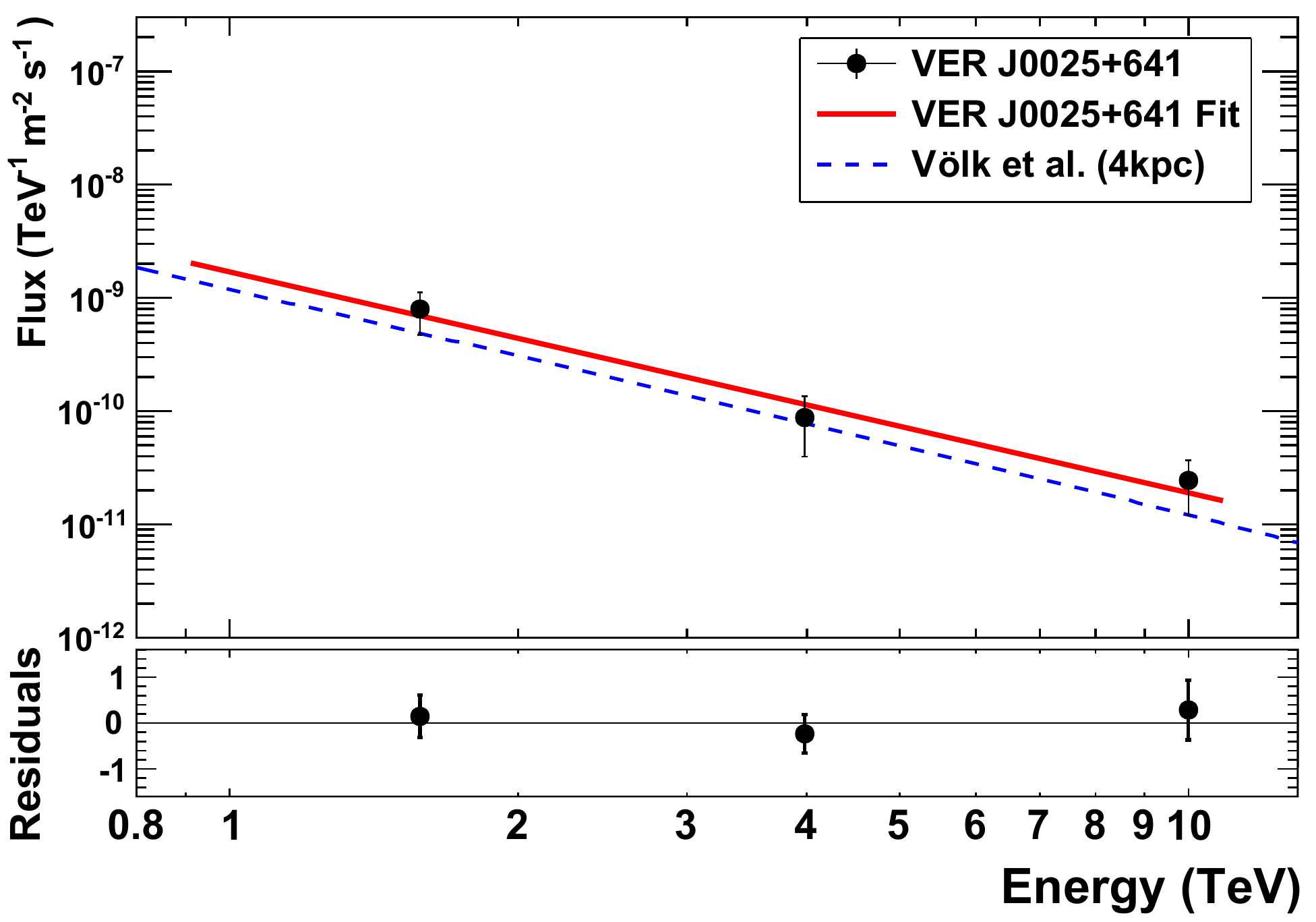}
\caption{\label{fig:spectrum} Differential gamma-ray photon spectrum of Tycho as measured by VERITAS.  The error bars represent $1\sigma$ statistical errors only.  The solid red line shows the results of a power-law fit to the VERITAS data.  The lower panel shows the residuals of the data from this fit.  The dashed blue line represents the hadronic model of \cite{2008A&A...483..529V} for gamma-ray emission from Tycho, scaled to 4~kpc.  The details of the model and of the analysis are discussed in the text.}
\end{figure}

The differential photon spectrum between 1 and 10~TeV is shown in Figure \ref{fig:spectrum}.  This spectrum is generated from the complete data set after quality selection.  The shape is consistent with a power law $dN/dE = C(E/3.42~\textrm{TeV})^{-\Gamma}$ with $\Gamma = 1.95 \pm 0.51_{stat} \pm 0.30_{sys}$ and $C = (1.55 \pm 0.43_{stat} \pm 0.47_{sys}) \times 10^{-14}$ cm$^{-2}$ s$^{-1}$ TeV$^{-1}$, where the systematic error on the flux is dominated by uncertainty in the energy scale.  The $\chi^2$ of the fit is 0.6 for 1 degree of freedom.  The integrated flux above 1 TeV is $(1.87 \pm 0.51_{stat}) \times 10^{-13}$ cm$^{-2}$ s$^{-1}$, about 0.9\% that of the steady Crab Nebula flux above the same energy.

\section{Discussion}

Figure \ref{fig:skymap} shows the TeV gamma-ray image of Tycho's SNR.  The color scale indicates the number of excess gamma-ray events in a region, using a squared integration radius of 0.015~deg$^2$ for the 2008/2009 data and 0.01~deg$^2$ for the 2009/2010 data.  The map has been smoothed with a Gaussian kernel of radius 0.06$^{\circ}$.  Overlaid on the image are X-ray contours from a Chandra ACIS exposure \citep[thin black lines;][]{2005ApJ...634..376W}.  A contour map of 115~GHz line emission associated with the molecular $^{12}$CO  (J=1-0) transition, from the 14 m telescope of the Five College Radio Astronomy Observatory (FCRAO), is shown in magenta \citep{1998ApJS..115..241H,2003AJ....125.3145T}.  Following the analysis of \cite{2004ApJ...605L.113L}, we have integrated over the velocity range -68~km~s$^{-1}$ to -50~km~s$^{-1}$, revealing a cloud possibly interacting with the northeast quadrant of the remnant.

As can be seen from Figure \ref{fig:skymap}, the peak of the gamma-ray emission is displaced somewhat to the northeast of the center of the remnant, towards the CO cloud. While this is provocative in the context of hadronically-induced gamma-ray emission \citep[see, e.g.,][]{1994A&A...285..645A}, the statistical significance of the displacement is weak.  Nevertheless, this is the general morphology expected in the case of a shock/cloud interaction leading to gamma-ray emission, which is seen in several other remnants \citep[see, e.g.,][]{2009ApJ...706L.270H}.  On the other hand, OH maser emission, a telltale sign of such interactions \citep{2002Sci...296.2350W} has not been detected from this remnant \citep{1996AJ....111.1651F}.  A catalog search within our error box reveals no likely gamma-ray emission candidates at other wavelengths, so we tentatively associate the source with Tycho.

\begin{figure}
\includegraphics[width=0.90\linewidth]{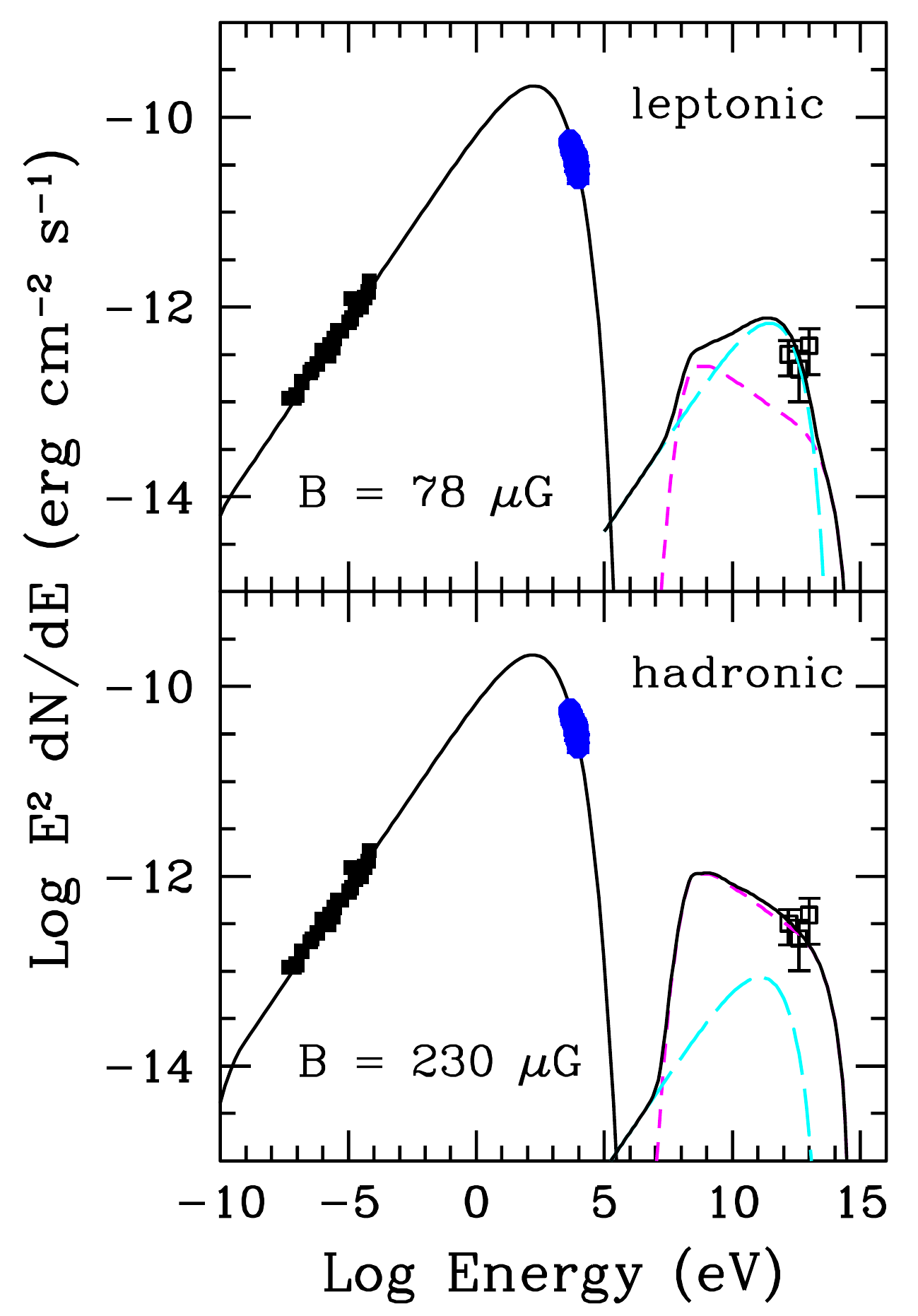}
\caption{\label{fig:models}Radio, (nonthermal) X-ray, and VHE gamma-ray emission from Tycho's SNR, along with models for the emission. The upper panel shows a lepton-dominated model while the lower panel shows a model dominated by hadrons. In each, the IC emission corresponds to the (cyan) long-dashed curve while the pion-decay emission corresponds to the (magenta) short-dashed curve. The solid curve at high energies is the sum of these components; at lower energies it corresponds to the synchrotron emission.  See text for discussion.}
\end{figure}


In Figure \ref{fig:models} we show simple model fits to the broadband spectrum of Tycho, generated assuming no influence from the molecular cloud.  Two versions are considered - one in which the TeV emission is dominated by leptonic (inverse Compton (IC) scattering; upper panel) processes, and one in which it is dominated by hadronic (pion decay; lower panel) processes \citep{2010ApJ...720..266S}. Here we have assumed particle spectra of the form:
\begin{equation}
\frac{dN}{dE} = A E^{-\alpha} e^{-(E/E_c)}
\end{equation}
with the spectral index $\alpha$ being fixed to the same value for the electrons and protons, but with the cutoff energy $E_c$ being allowed to differ, as expected for loss-dominated distributions. We note the crucial point that, in both the leptonic and hadronic models, the total relativistic particle energy is dominated by hadrons, and it represents a significant fraction of the total supernova kinetic energy, estimated to be $1.2 \times 10^{51}$ ergs \citep{2006ApJ...645.1373B}.  Indeed, in both cases, the energy density of hadronic cosmic rays is likely sufficient to modify the supernova shock dynamics, supporting the conclusions of \cite{2005ApJ...634..376W}.  The radio spectrum requires $\alpha \approx 2.2$, and we have assumed an ambient density $n_0 = 0.2 {\rm\ cm}^{-3}$, based on upper limits from X-ray measurements \citep{2010ApJ...709.1387K}.  The distance is set to 4~kpc.

The radio data in Figure \ref{fig:models} are compiled from \cite{1992ApJ...399L..75R}. The X-ray data (shown in blue) represent the unfolded spectrum between $\sim 4$ and $ 10$~keV extracted from a deep {\it Suzaku} observation of Tycho. Features from Fe-K and weaker line emission have been removed. The TeV data points are from the VERITAS results reported here.

For the lepton-dominated scenario, we find the normalization and magnetic field required to reproduce the radio and X-ray emission by synchrotron radiation.  The gamma-ray emission is then generated by IC scattering of cosmic microwave background photons. (The impact of adding an additional - even maximal - contribution from the dust-IR emission \citep{2001A&A...373..281D} was investigated and found to make only a small difference in the overall IC spectrum.) The associated magnetic field is $\sim 80 \mu{\rm G}$, with a $\sim 15\%$ uncertainty.  Assuming an electron-to-proton number ratio $\kappa_{ep} = 10^{-2}$, as observed in the local cosmic-ray population, the total particle energy (dominated by protons) is $1.8 \times 10^{50}{\rm\ erg}$, and the associated gamma-ray emission from $\pi^0$ decay is negligible in the TeV band, as shown in the upper panel of Figure \ref{fig:models}.  The magnetic field required for this model is somewhat lower than the conventionally accepted $\sim 200-300 \mu$G value that many derive \citep[e.g.,][]{2006AdSpR..37.1902B,2008A&A...483..529V,2010A&A...515A..90M}, though it is within the range found by \cite{2005ApJ...634..376W}.

For the hadron-dominated scenario, we adjust the normalization of the proton spectrum to reproduce the observed TeV emission through $\pi^0$-decay (Figure \ref{fig:models}, lower panel). The normalization of the electron spectrum is then reduced to a level at which the hadron-induced gamma-ray emission dominates, yielding a $\kappa_{ep} \approx 4 \times 10^{-4}$.  The associated magnetic field required to reproduce the synchrotron emission is $\sim 230 \mu{\rm G}$.  This value can be considered a lower limit, since a smaller $\kappa_{ep}$ will require a larger magnetic field.

For this model, we find a total particle energy of $\sim 8 \times 10^{50}{\rm\ erg}$.  While this is perhaps uncomfortably large, it is based on emission involving only the mean gas density of the remnant.  Any possible contribution of target material from the cloud would reduce the energy requirement.  This could also be achieved if we loosen the assumption of a single power-law particle spectrum.




Our hadronic model is broadly consistent with that of \cite{2008A&A...483..529V}, who employ a nonlinear kinetic particle acceleration model to derive a hadronically-induced flux as a function of distance.  As shown in Figure \ref{fig:spectrum}, with a source distance of $4$~kpc, an ambient density of $\sim0.2$~cm$^{-3}$, and a magnetic field of $\sim350\mu{\rm G}$, their model provides a reasonable fit to our data.  However, by scaling the flux to improve the fit, a distance estimate of $3.8$~kpc can be obtained.

Finally, it is worth noting that the lowest magnetic field allowable in either of our models is $\sim 80 \mu{\rm G}$.  This value is not only well above the typical $\sim 3\mu{\rm G}$ fields of the interstellar medium, it is significantly higher than expectations from shock-compression of that field in moderate-density environments \citep[see, e.g.,][]{2008ApJ...673L..47E}.  Hence, our detection of gamma rays from Tycho's SNR represents additional evidence for magnetic field amplification in this remnant.  While this conclusion is dependent on the validity of our one-zone emission approximation, it is not subject to large systematic uncertainties arising from the choice of electron spectrum, since, for a $\sim 80 \mu{\rm G}$ field, the \textit{Suzaku} and VERITAS data probe the same underlying electron energies.

\section{Conclusion}

VERITAS has detected weak TeV gamma-ray emission from the SNR G120.1+1.4, also known as Tycho's supernova remnant.  The total flux from the remnant above 1 TeV is $\sim0.9\%$ that of the Crab Nebula, making it one of the weakest sources yet detected in TeV gamma rays, and only the second confirmed Type Ia SNR gamma-ray emitter. SN1006, another Type Ia remnant, was observed by H.E.S.S. to have two distinct regions of weak ($\sim 1\%$ Crab) gamma-ray emission, one at the northeastern edge of the $\sim 0.5^\circ$ diameter remnant and one at the southwestern edge \citep{2010A&A...516A..62A}. Both of these regions are highly correlated with non-thermal X-ray emission and have spectra that are compatible with power-laws of index $\Gamma \sim 2.3$.  For SN1006, neither a leptonic nor a hadronic origin for gamma-ray emission can be firmly eliminated.



The photon spectrum of Tycho can be described by a power-law with differential index $1.95 \pm 0.51_{stat} \pm 0.30_{sys}$ and is compatible with the hadronic model of \cite{2008A&A...483..529V}, scaled for a distance of $3.8$~kpc.  We present another hadronic model which can also reproduce the data, although it requires a large energy content of cosmic rays.  A leptonic external IC model also provides a tolerable fit to the data, though it requires a magnetic field somewhat lower than generally is accepted for Tycho.  Notably, the lowest magnetic field allowed in either model is $\sim 80 \mu$G, which may be interpreted as evidence for magnetic field amplification.  The morphology of the emission is compatible with a point source, with the peak of the emission possibly offset from the center of the remnant.  Detailed spectral and spatial studies will be possible with a deeper exposure.

\acknowledgements
This research is supported by grants from the U.S. Department of Energy, the U.S. National Science Foundation and the Smithsonian Institution, by the Natural Sciences and Engineering Research Council (NSERC) in Canada, by Science Foundation Ireland (SFI 10/RFP/AST2748) and by the Science and Technology Facilities Council in the UK. The research presented in this paper has used data from the Canadian Galactic Plane Survey, a Canadian project with international partners, supported by NSERC. D.B. Saxon acknowledges the NASA Delaware Space Grant Program for its support of this research. J.P. Hughes acknowledges support from NASA grant NNX08AZ86G to Rutgers University.

{\it Facilities:} \facility{FLWO:VERITAS}.


\begin{thebibliography}{50}
\expandafter\ifx\csname natexlab\endcsname\relax\def\natexlab#1{#1}\fi

\bibitem[{{Abdo} {et~al.}(2010){Abdo}, {Ackermann}, {Ajello}, {Allafort},
  {Antolini}, {Atwood}, {Axelsson}, {Baldini}, {Ballet}, {Barbiellini}, \&
  et~al.}]{2010ApJS..188..405A}
{Abdo}, A.~A. {et~al.} 2010, \apjs, 188, 405, 1002.2280

\bibitem[{{Acciari} {et~al.}(2008){Acciari}, {Beilicke}, {Blaylock},
  {Bradbury}, {Buckley}, {Bugaev}, {Butt}, {Byrum}, {Celik}, {Cesarini},
  {Ciupik}, {Chow}, {Cogan}, {Colin}, {Cui}, {Daniel}, {Duke}, {Ergin},
  {Falcone}, {Fegan}, {Finley}, {Fortin}, {Fortson}, {Gall}, {Gibbs},
  {Gillanders}, {Grube}, {Guenette}, {Hanna}, {Hays}, {Holder}, {Horan},
  {Hughes}, {Hui}, {Humensky}, {Kaaret}, {Kieda}, {Kildea}, {Konopelko},
  {Krawczynski}, {Krennrich}, {Lang}, {LeBohec}, {Lee}, {Maier}, {McCann},
  {McCutcheon}, {Millis}, {Moriarty}, {Mukherjee}, {Nagai}, {Ong}, {Pandel},
  {Perkins}, {Pizlo}, {Pohl}, {Quinn}, {Ragan}, {Reynolds}, {Rose},
  {Schroedter}, {Sembroski}, {Smith}, {Steele}, {Swordy}, {Toner}, {Valcarcel},
  {Vassiliev}, {Wagner}, {Wakely}, {Ward}, {Weekes}, {Weinstein}, {White},
  {Williams}, {Wissel}, {Wood}, \& {Zitzer}}]{2008ApJ...679.1427A}
{Acciari}, V.~A. {et~al.} 2008, \apj, 679, 1427, 0802.2363

\bibitem[{{Acero} {et~al.}(2010){Acero}, {Aharonian}, {Akhperjanian}, {Anton},
  {Barres de Almeida}, {Bazer-Bachi}, {Becherini}, {Behera}, {Beilicke},
  {Bernl{\"o}hr}, {Bochow}, {Boisson}, {Bolmont}, {Borrel}, {Brucker}, {Brun},
  {Brun}, {B{\"u}hler}, {Bulik}, {B{\"u}sching}, {Boutelier}, {Chadwick},
  {Charbonnier}, {Chaves}, {Cheesebrough}, {Conrad}, {Chounet}, {Clapson},
  {Coignet}, {Dalton}, {Daniel}, {Davids}, {Degrange}, {Deil}, {Dickinson},
  {Djannati-Ata{\"i}}, {Domainko}, {O'C.~Drury}, {Dubois}, {Dubus}, {Dyks},
  {Dyrda}, {Egberts}, {Eger}, {Espigat}, {Fallon}, {Farnier}, {Fegan},
  {Feinstein}, {Fiasson}, {F{\"o}rster}, {Fontaine}, {F{\"u}{\ss}ling},
  {Gabici}, {Gallant}, {G{\'e}rard}, {Gerbig}, {Giebels}, {Glicenstein},
  {Gl{\"u}ck}, {Goret}, {G{\"o}ring}, {Hauser}, {Hauser}, {Heinz},
  {Heinzelmann}, {Henri}, {Hermann}, {Hinton}, {Hoffmann}, {Hofmann},
  {Hofverberg}, {Holleran}, {Hoppe}, {Horns}, {Jacholkowska}, {de Jager},
  {Jahn}, {Jung}, {Katarzy{\'n}ski}, {Katz}, {Kaufmann}, {Kerschhaggl},
  {Khangulyan}, {Kh{\'e}lifi}, {Keogh}, {Klochkov}, {Klu{\'z}niak}, {Kneiske},
  {Komin}, {Kosack}, {Kossakowski}, {Lamanna}, {Lemoine-Goumard}, {Lenain},
  {Lohse}, {Marandon}, {Marcowith}, {Masbou}, {Maurin}, {McComb}, {Medina},
  {M{\'e}hault}, {Moderski}, {Moulin}, {Naumann-Godo}, {de Naurois}, {Nedbal},
  {Nekrassov}, {Nicholas}, {Niemiec}, {Nolan}, {Ohm}, {Olive}, {de O{\~n}a
  Wilhelmi}, {Orford}, {Ostrowski}, {Panter}, {Paz Arribas}, {Pedaletti},
  {Pelletier}, {Petrucci}, {Pita}, {P{\"u}hlhofer}, {Punch}, {Quirrenbach},
  {Raubenheimer}, {Raue}, {Rayner}, {Reimer}, {Renaud}, {de Los Reyes},
  {Rieger}, {Ripken}, {Rob}, {Rosier-Lees}, {Rowell}, {Rudak}, {Rulten},
  {Ruppel}, {Ryde}, {Sahakian}, {Santangelo}, {Schlickeiser}, {Sch{\"o}ck},
  {Sch{\"o}nwald}, {Schwanke}, {Schwarzburg}, {Schwemmer}, {Shalchi}, {Sushch},
  {Sikora}, {Skilton}, {Sol}, {Stawarz}, {Steenkamp}, {Stegmann}, {Stinzing},
  {Superina}, {Szostek}, {Tam}, {Tavernet}, {Terrier}, {Tibolla}, {Tluczykont},
  {van Eldik}, {Vasileiadis}, {Venter}, {Venter}, {Vialle}, {Vincent}, {Vink},
  {Vivier}, {V{\"o}lk}, {Volpe}, {Vorobiov}, {Wagner}, {Ward}, {Zdziarski},
  {Zech}, \& {HESS Collaboration}}]{2010A&A...516A..62A}
{Acero}, F. {et~al.} 2010, \aap, 516, A62+, 1004.2124

\bibitem[{{Aharonian} {et~al.}(2006){Aharonian}, {Akhperjanian}, {Bazer-Bachi},
  {Beilicke}, {Benbow}, {Berge}, {Bernl{\"o}hr}, {Boisson}, {Bolz}, {Borrel},
  {Braun}, {Breitling}, {Brown}, {Chadwick}, {Chounet}, {Cornils},
  {Costamante}, {Degrange}, {Dickinson}, {Djannati-Ata{\"i}}, {Drury}, {Dubus},
  {Emmanoulopoulos}, {Espigat}, {Feinstein}, {Fontaine}, {Fuchs}, {Funk},
  {Gallant}, {Giebels}, {Gillessen}, {Glicenstein}, {Goret}, {Hadjichristidis},
  {Hauser}, {Heinzelmann}, {Henri}, {Hermann}, {Hinton}, {Hofmann}, {Holleran},
  {Horns}, {Jacholkowska}, {de Jager}, {Kh{\'e}lifi}, {Komin}, {Konopelko},
  {Latham}, {Le Gallou}, {Lemi{\`e}re}, {Lemoine-Goumard}, {Leroy}, {Lohse},
  {Martin}, {Martineau-Huynh}, {Marcowith}, {Masterson}, {McComb}, {de
  Naurois}, {Nolan}, {Noutsos}, {Orford}, {Osborne}, {Ouchrif}, {Panter},
  {Pelletier}, {Pita}, {P{\"u}hlhofer}, {Punch}, {Raubenheimer}, {Raue},
  {Raux}, {Rayner}, {Reimer}, {Reimer}, {Ripken}, {Rob}, {Rolland}, {Rowell},
  {Sahakian}, {Saug{\'e}}, {Schlenker}, {Schlickeiser}, {Schuster}, {Schwanke},
  {Siewert}, {Sol}, {Spangler}, {Steenkamp}, {Stegmann}, {Tavernet}, {Terrier},
  {Th{\'e}oret}, {Tluczykont}, {Vasileiadis}, {Venter}, {Vincent}, {V{\"o}lk},
  \& {Wagner}}]{2006ApJ...636..777A}
{Aharonian}, F. {et~al.} 2006, \apj, 636, 777, arXiv:astro-ph/0510397

\bibitem[{{Aharonian} {et~al.}(2001){Aharonian}, {Akhperjanian}, {Barrio},
  {Bernl{\"o}hr}, {B{\"o}rst}, {Bojahr}, {Bolz}, {Contreras}, {Cortina},
  {Denninghoff}, {Fonseca}, {Gonzalez}, {G{\"o}tting}, {Heinzelmann},
  {Hermann}, {Heusler}, {Hofmann}, {Horns}, {Ibarra}, {Jung}, {Kankanyan},
  {Kestel}, {Kettler}, {Kohnle}, {Konopelko}, {Kornmeyer}, {Kranich},
  {Krawczynski}, {Lampeitl}, {Lorenz}, {Lucarelli}, {Magnussen}, {Mang},
  {Meyer}, {Mirzoyan}, {Moralejo}, {Padilla}, {Panter}, {Plaga},
  {Plyasheshnikov}, {Prahl}, {P{\"u}hlhofer}, {Rauterberg}, {R{\"o}hring},
  {Rhode}, {Rowell}, {Sahakian}, {Samorski}, {Schilling}, {Schr{\"o}der},
  {Stamm}, {Tluczykont}, {V{\"o}lk}, {Wiedner}, \&
  {Wittek}}]{2001A&A...373..292A}
{Aharonian}, F.~A. {et~al.} 2001, \aap, 373, 292, arXiv:astro-ph/0107044

\bibitem[{{Aharonian} {et~al.}(1994){Aharonian}, {Drury}, \&
  {Voelk}}]{1994A&A...285..645A}
{Aharonian}, F.~A., {Drury}, L.~O., \& {Voelk}, H.~J. 1994, \aap, 285, 645

\bibitem[{{Badenes} {et~al.}(2006){Badenes}, {Borkowski}, {Hughes}, {Hwang}, \&
  {Bravo}}]{2006ApJ...645.1373B}
{Badenes}, C., {Borkowski}, K.~J., {Hughes}, J.~P., {Hwang}, U., \& {Bravo}, E.
  2006, \apj, 645, 1373, arXiv:astro-ph/0511140

\bibitem[{{Ballet}(2006)}]{2006AdSpR..37.1902B}
{Ballet}, J. 2006, Advances in Space Research, 37, 1902, arXiv:astro-ph/0503309

\bibitem[{{Bamba} {et~al.}(2005){Bamba}, {Yamazaki}, {Yoshida}, {Terasawa}, \&
  {Koyama}}]{2005ApJ...621..793B}
{Bamba}, A., {Yamazaki}, R., {Yoshida}, T., {Terasawa}, T., \& {Koyama}, K.
  2005, \apj, 621, 793, arXiv:astro-ph/0411326

\bibitem[{{Berge} {et~al.}(2007){Berge}, {Funk}, \&
  {Hinton}}]{2007A&A...466.1219B}
{Berge}, D., {Funk}, S., \& {Hinton}, J. 2007, \aap, 466, 1219,
  arXiv:astro-ph/0610959

\bibitem[{{Buckley} {et~al.}(1998){Buckley}, {Akerlof}, {Carter-Lewis},
  {Catanese}, {Cawley}, {Connaughton}, {Fegan}, {Finley}, {Gaidos}, {Hillas},
  {Krennrich}, {Lamb}, {Lessard}, {McEnery}, {Mohanty}, {Quinn}, {Rodgers},
  {Rose}, {Rovero}, {Schubnell}, {Sembroski}, {Srinivasan}, {Weekes}, \&
  {Zweerink}}]{1998A&A...329..639B}
{Buckley}, J.~H. {et~al.} 1998, \aap, 329, 639

\bibitem[{{Cai} {et~al.}(2009){Cai}, {Yang}, \& {Lu}}]{2009ChA&A..33..393C}
{Cai}, Z., {Yang}, J., \& {Lu}, D. 2009, Chinese Astronomy and Astrophysics,
  33, 393

\bibitem[{{Carmona} {et~al.}(2009){Carmona}, {Costado}, {Font}, {Zapatero}, \&
  {for the MAGIC Collaboration}}]{2009arXiv0907.1009C}
{Carmona}, E., {Costado}, M.~T., {Font}, L., {Zapatero}, J., \& {for the MAGIC
  Collaboration}. 2009, ArXiv e-prints, 0907.1009

\bibitem[{{Chevalier} {et~al.}(1980){Chevalier}, {Kirshner}, \&
  {Raymond}}]{1980ApJ...235..186C}
{Chevalier}, R.~A., {Kirshner}, R.~P., \& {Raymond}, J.~C. 1980, \apj, 235, 186

\bibitem[{{Dickel} {et~al.}(1991){Dickel}, {van Breugel}, \&
  {Strom}}]{1991AJ....101.2151D}
{Dickel}, J.~R., {van Breugel}, W.~J.~M., \& {Strom}, R.~G. 1991, \aj, 101,
  2151

\bibitem[{{Douvion} {et~al.}(2001){Douvion}, {Lagage}, {Cesarsky}, \&
  {Dwek}}]{2001A&A...373..281D}
{Douvion}, T., {Lagage}, P.~O., {Cesarsky}, C.~J., \& {Dwek}, E. 2001, \aap,
  373, 281

\bibitem[{{Ellison} \& {Vladimirov}(2008)}]{2008ApJ...673L..47E}
{Ellison}, D.~C., \& {Vladimirov}, A. 2008, \apjl, 673, L47, 0711.4389

\bibitem[{{Frail} {et~al.}(1996){Frail}, {Goss}, {Reynoso}, {Giacani}, {Green},
  \& {Otrupcek}}]{1996AJ....111.1651F}
{Frail}, D.~A., {Goss}, W.~M., {Reynoso}, E.~M., {Giacani}, E.~B., {Green},
  A.~J., \& {Otrupcek}, R. 1996, \aj, 111, 1651

\bibitem[{{Hartman} {et~al.}(1999){Hartman}, {Bertsch}, {Bloom}, {Chen},
  {Deines-Jones}, {Esposito}, {Fichtel}, {Friedlander}, {Hunter}, {McDonald},
  {Sreekumar}, {Thompson}, {Jones}, {Lin}, {Michelson}, {Nolan}, {Tompkins},
  {Kanbach}, {Mayer-Hasselwander}, {M{\"u}cke}, {Pohl}, {Reimer}, {Kniffen},
  {Schneid}, {von Montigny}, {Mukherjee}, \& {Dingus}}]{1999ApJS..123...79H}
{Hartman}, R.~C. {et~al.} 1999, \apjs, 123, 79

\bibitem[{{Hayato} {et~al.}(2010){Hayato}, {Yamaguchi}, {Tamagawa}, {Katsuda},
  {Hwang}, {Hughes}, {Ozawa}, {Bamba}, {Kinugasa}, {Terada}, {Furuzawa},
  {Kunieda}, \& {Makishima}}]{2010arXiv1009.6031H}
{Hayato}, A. {et~al.} 2010, ArXiv e-prints, 1009.6031

\bibitem[{{Helder} {et~al.}(2009){Helder}, {Vink}, {Bassa}, {Bamba}, {Bleeker},
  {Funk}, {Ghavamian}, {van der Heyden}, {Verbunt}, \&
  {Yamazaki}}]{2009Sci...325..719H}
{Helder}, E.~A. {et~al.} 2009, Science, 325, 719, 0906.4553

\bibitem[{{Hewitt} {et~al.}(2009){Hewitt}, {Yusef-Zadeh}, \&
  {Wardle}}]{2009ApJ...706L.270H}
{Hewitt}, J.~W., {Yusef-Zadeh}, F., \& {Wardle}, M. 2009, \apjl, 706, L270,
  0909.2827

\bibitem[{{Heyer} {et~al.}(1998){Heyer}, {Brunt}, {Snell}, {Howe}, {Schloerb},
  \& {Carpenter}}]{1998ApJS..115..241H}
{Heyer}, M.~H., {Brunt}, C., {Snell}, R.~L., {Howe}, J.~E., {Schloerb}, F.~P.,
  \& {Carpenter}, J.~M. 1998, \apjs, 115, 241

\bibitem[{{Hwang} {et~al.}(2002){Hwang}, {Petre}, {Szymkowiak}, \&
  {Holt}}]{2002JApA...23...81H}
{Hwang}, U., {Petre}, R., {Szymkowiak}, A.~E., \& {Holt}, S.~S. 2002, Journal
  of Astrophysics and Astronomy, 23, 81

\bibitem[{{Kaaret} {et~al.}(2009){Kaaret}, {Butt}, {Digel}, {Funk}, {Halzen},
  {Hanna}, {Heinz}, {Gyuk}, {Hays}, {LeBohec}, {Meszaros}, {Moskalenko},
  {Mukherjee}, {Ong}, {Pohl}, {Romani}, {Sinnis}, {Slane}, \&
  {Wakely}}]{2009astro2010S.145K}
{Kaaret}, P. {et~al.} 2009, in ArXiv Astrophysics e-prints, Vol. 2010,
  astro2010: The Astronomy and Astrophysics Decadal Survey, 145--+

\bibitem[{{Kamper} \& {van den Bergh}(1978)}]{1978ApJ...224..851K}
{Kamper}, K.~W., \& {van den Bergh}, S. 1978, \apj, 224, 851

\bibitem[{{Katsuda} {et~al.}(2010){Katsuda}, {Petre}, {Hughes}, {Hwang},
  {Yamaguchi}, {Hayato}, {Mori}, \& {Tsunemi}}]{2010ApJ...709.1387K}
{Katsuda}, S., {Petre}, R., {Hughes}, J.~P., {Hwang}, U., {Yamaguchi}, H.,
  {Hayato}, A., {Mori}, K., \& {Tsunemi}, H. 2010, \apj, 709, 1387, 1001.2484

\bibitem[{{Katz-Stone} {et~al.}(2000){Katz-Stone}, {Kassim}, {Lazio}, \&
  {O'Donnell}}]{2000ApJ...529..453K}
{Katz-Stone}, D.~M., {Kassim}, N.~E., {Lazio}, T.~J.~W., \& {O'Donnell}, R.
  2000, \apj, 529, 453

\bibitem[{{Kothes} {et~al.}(2006){Kothes}, {Fedotov}, {Foster}, \&
  {Uyan{\i}ker}}]{2006A&A...457.1081K}
{Kothes}, R., {Fedotov}, K., {Foster}, T.~J., \& {Uyan{\i}ker}, B. 2006, \aap,
  457, 1081

\bibitem[{{Krause} {et~al.}(2008){Krause}, {Tanaka}, {Usuda}, {Hattori},
  {Goto}, {Birkmann}, \& {Nomoto}}]{2008Natur.456..617K}
{Krause}, O., {Tanaka}, M., {Usuda}, T., {Hattori}, T., {Goto}, M., {Birkmann},
  S., \& {Nomoto}, K. 2008, \nat, 456, 617, 0810.5106

\bibitem[{{Krawczynski} {et~al.}(2006){Krawczynski}, {Carter-Lewis}, {Duke},
  {Holder}, {Maier}, {Le Bohec}, \& {Sembroski}}]{2006APh....25..380K}
{Krawczynski}, H., {Carter-Lewis}, D.~A., {Duke}, C., {Holder}, J., {Maier},
  G., {Le Bohec}, S., \& {Sembroski}, G. 2006, Astroparticle Physics, 25, 380,
  arXiv:astro-ph/0604508

\bibitem[{{Lee} {et~al.}(2004){Lee}, {Koo}, \&
  {Tatematsu}}]{2004ApJ...605L.113L}
{Lee}, J., {Koo}, B., \& {Tatematsu}, K. 2004, \apjl, 605, L113,
  arXiv:astro-ph/0403274

\bibitem[{{Lee} {et~al.}(2010){Lee}, {Raymond}, {Park}, {Blair}, {Ghavamian},
  {Winkler}, \& {Korreck}}]{2010ApJ...715L.146L}
{Lee}, J., {Raymond}, J.~C., {Park}, S., {Blair}, W.~P., {Ghavamian}, P.,
  {Winkler}, P.~F., \& {Korreck}, K. 2010, \apjl, 715, L146, 1005.3296

\bibitem[{{Marcowith} \& {Casse}(2010)}]{2010A&A...515A..90M}
{Marcowith}, A., \& {Casse}, F. 2010, \aap, 515, A90+, 1001.2111

\bibitem[{{Perkins} {et~al.}(2009){Perkins}, {Maier}, \& {The VERITAS
  Collaboration}}]{2009arXiv0912.3841P}
{Perkins}, J.~S., {Maier}, G., \& {The VERITAS Collaboration}. 2009, ArXiv
  e-prints, 0912.3841

\bibitem[{{Reynolds} \& {Ellison}(1992)}]{1992ApJ...399L..75R}
{Reynolds}, S.~P., \& {Ellison}, D.~C. 1992, \apjl, 399, L75

\bibitem[{{Reynoso} {et~al.}(1997){Reynoso}, {Moffett}, {Goss}, {Dubner},
  {Dickel}, {Reynolds}, \& {Giacani}}]{1997ApJ...491..816R}
{Reynoso}, E.~M., {Moffett}, D.~A., {Goss}, W.~M., {Dubner}, G.~M., {Dickel},
  J.~R., {Reynolds}, S.~P., \& {Giacani}, E.~B. 1997, \apj, 491, 816

\bibitem[{{Reynoso} {et~al.}(1999){Reynoso}, {Vel{\'a}zquez}, {Dubner}, \&
  {Goss}}]{1999AJ....117.1827R}
{Reynoso}, E.~M., {Vel{\'a}zquez}, P.~F., {Dubner}, G.~M., \& {Goss}, W.~M.
  1999, \aj, 117, 1827

\bibitem[{{Ruiz-Lapuente}(2004)}]{2004ApJ...612..357R}
{Ruiz-Lapuente}, P. 2004, \apj, 612, 357, arXiv:astro-ph/0309009

\bibitem[{{Schwarz} {et~al.}(1995){Schwarz}, {Goss}, {Kalberla}, \&
  {Benaglia}}]{1995A&A...299..193S}
{Schwarz}, U.~J., {Goss}, W.~M., {Kalberla}, P.~M., \& {Benaglia}, P. 1995,
  \aap, 299, 193

\bibitem[{{Slane} {et~al.}(2010){Slane}, {Castro}, {Funk}, {Uchiyama},
  {Lemiere}, {Gelfand}, \& {Lemoine-Goumard}}]{2010ApJ...720..266S}
{Slane}, P., {Castro}, D., {Funk}, S., {Uchiyama}, Y., {Lemiere}, A.,
  {Gelfand}, J.~D., \& {Lemoine-Goumard}, M. 2010, \apj, 720, 266

\bibitem[{{Smith} {et~al.}(1991){Smith}, {Kirshner}, {Blair}, \&
  {Winkler}}]{1991ApJ...375..652S}
{Smith}, R.~C., {Kirshner}, R.~P., {Blair}, W.~P., \& {Winkler}, P.~F. 1991,
  \apj, 375, 652

\bibitem[{{Smith} {et~al.}(1992){Smith}, {Kirshner}, {Blair}, \&
  {Winkler}}]{1992ApJ...384..665S}
------. 1992, \apj, 384, 665

\bibitem[{{Strom}(1988)}]{1988MNRAS.230..331S}
{Strom}, R.~G. 1988, \mnras, 230, 331

\bibitem[{{Stroman} \& {Pohl}(2009)}]{2009ApJ...696.1864S}
{Stroman}, W., \& {Pohl}, M. 2009, \apj, 696, 1864, 0902.1701

\bibitem[{{Tamagawa} {et~al.}(2009){Tamagawa}, {Hayato}, {Nakamura}, {Terada},
  {Bamba}, {Hiraga}, {Hughes}, {Hwang}, {Kataoka}, {Kinugasa}, {Kunieda},
  {Tanaka}, {Tsunemi}, {Ueno}, {Holt}, {Kokubun}, {Miyata}, {Szymkowiak},
  {Takahashi}, {Tamura}, {Ueno}, \& {Makishima}}]{2009PASJ...61S.167T}
{Tamagawa}, T. {et~al.} 2009, \pasj, 61, 167, 0805.3377

\bibitem[{{Taylor} {et~al.}(2003){Taylor}, {Gibson}, {Peracaula}, {Martin},
  {Landecker}, {Brunt}, {Dewdney}, {Dougherty}, {Gray}, {Higgs}, {Kerton},
  {Knee}, {Kothes}, {Purton}, {Uyaniker}, {Wallace}, {Willis}, \&
  {Durand}}]{2003AJ....125.3145T}
{Taylor}, A.~R. {et~al.} 2003, \aj, 125, 3145

\bibitem[{{V{\"o}lk} {et~al.}(2008){V{\"o}lk}, {Berezhko}, \&
  {Ksenofontov}}]{2008A&A...483..529V}
{V{\"o}lk}, H.~J., {Berezhko}, E.~G., \& {Ksenofontov}, L.~T. 2008, \aap, 483,
  529, 0803.1403

\bibitem[{{Wardle} \& {Yusef-Zadeh}(2002)}]{2002Sci...296.2350W}
{Wardle}, M., \& {Yusef-Zadeh}, F. 2002, Science, 296, 2350

\bibitem[{{Warren} {et~al.}(2005){Warren}, {Hughes}, {Badenes}, {Ghavamian},
  {McKee}, {Moffett}, {Plucinsky}, {Rakowski}, {Reynoso}, \&
  {Slane}}]{2005ApJ...634..376W}
{Warren}, J.~S. {et~al.} 2005, \apj, 634, 376, arXiv:astro-ph/0507478

\end{thebibliography}
\end{document}